\documentclass[aps,prb,twocolumn,showpacs,preprintnumbers,amsmath,amssymb,superscriptaddress,floatfix]{revtex4}
\usepackage{graphicx}
\usepackage{dcolumn}
\usepackage{bm}
\usepackage{amsmath,amssymb}
\newcommand{\ee}{\mathrm e}

\newcommand{\br}{\mathbf{r}}
\newcommand{\bq}{\mathbf{q}}

\newcommand{\bk}{\mathbf{k}}
\newcommand{\bkp}{\mathbf{k'}}

\newcommand{\op}[1]{\hat{#1}}
\newcommand{\ad}{a^\dag}
\newcommand{\dd}{\mathrm{d}}

\newcommand{\bd}{b^\dag}
\begin{document}
\renewcommand{\thefigure}{\arabic{figure}}
\title{Stability of Sarma phases in density imbalanced electron-hole 
bilayer systems}
\author{A.L. Suba{\c s}{\i}}
\altaffiliation[Current address: ]{Faculty of Engineering and 
Natural Sciences, Sabanci University, Tuzla, 34956 Istanbul, Turkey}
\affiliation{Department of Physics, Bilkent University, Bilkent, 
Ankara 06800, Turkey}
\author{P. Pieri}
\affiliation{Dipartimento di Fisica, Scuola di Scienze e Tecnologie, Universit\`{a} di Camerino, 
I-62032 Camerino, Italy}
\author{G. Senatore}
\affiliation{CNR-INFM DEMOCRITOS National Simulation Center, Trieste, Italy}
\affiliation{Dipartimento di Fisica, Universit\`a di Trieste, 
Strada Costiera 11, I-34151 Trieste,Italy}
\author{B. Tanatar}
\affiliation{Department of Physics, Bilkent University, Bilkent, Ankara 
06800, Turkey}

\begin{abstract}
We study excitonic condensation in an electron-hole bilayer system 
with unequal layer densities at zero temperature. Using mean-field 
theory we solve the BCS gap equations numerically and investigate 
the effects of intra-layer interactions. We analyze the stability 
of the Sarma phase with $\bk,-\bk$ pairing by calculating the 
superfluid mass density and also by checking the 
compressibility matrix. 
We find that with bare Coulomb interactions 
the superfluid density is always positive in the Sarma phase, 
due to a peculiar momentum structure of the gap function originating 
from the singular behavior of the Coulomb potential at zero 
momentum and the presence of a sharp Fermi surface. 
Introducing a simple model for screening, 
we find that the superfluid density becomes negative in some regions of the 
phase diagram, corresponding to an instability towards a  
Fulde-Ferrel-Larkin-Ovchinnikov (FFLO) type superfluid phase. 
Thus, intra-layer interaction and screening together can lead 
to a rich phase diagram in the BCS-BEC crossover regime
in electron-hole bilayer systems.
\end{abstract}
\maketitle

\section{Introduction}

Recent advances in the trapping and cooling down to degeneracy
of ultracold Fermi gases have revived interest in the ground state 
phases of these systems.\cite{gre03,joc03,zwi03,chi04,reg04,
zwi05,zwi06,par06,sch07,ste08,jor08,jo09,blo08,trento} 
In a two-component Fermi system with equal densities attractive 
interactions between different species lead to Bardeen-Cooper-Schrieffer 
(BCS) pairing in the weak coupling limit and Bose-Einstein condensation 
(BEC) in the strong coupling limit.\cite{sademelo} 
When the densities are imbalanced more
exotic phases are expected to follow, such as the Sarma
phase\cite{sarma} with zero center-of-mass momentum and
Fulde-Ferrell-Larkin-Ovchinnikov (FFLO) phase\cite{ff,lo} with finite
center-of-mass momentum. There is a growing literature
on the possible phases of two-component Fermi gases with population 
and mass imbalance.
\cite{liu03,wu03,bed03,for05,car05,she06,che06,pie06,isk06,pao06,
chi06,gub06,nik07,par07,isk07,pil08,con08,sha08,ber09,he09,
che09,bau09,des09} 
The experimental efforts in ultracold
Fermi gases are in their beginning stage and so far only phase
separation between a superfluid and a normal phase has been
observed.\cite{zwi06,par06,shi06,par06b} 

Semiconducting electron-hole bilayer systems offer another
realization of a two-component Fermi system with which the exotic 
phases can be studied. Formation of excitons between spatially
separated electrons and holes and their subsequent condensation
have long been predicted\cite{lozovik,shevchenko} and arguably observed
nearly 30 years later experimentally.\cite{butov}
The phase diagram of symmetric electron-hole bilayer systems (equal 
mass and layer density) is most reliably calculated by quantum Monte
Carlo simulations. \cite{littlewood,senatore}
Recent success in fabricating closely spaced semiconducting
electron-hole bilayer structures\cite{keogh_apl,seamons_apl} 
and the ability to control the densities of individual layers 
make the investigation of Sarma and FFLO phases very timely. 
In fact, experiments supporting evidence of a transition from
the Fermi liquid phase to an excitonic condensate have been
recently reported through Coulomb drag
measurements\cite{croxall_prl,seamons_prl} in such structures.

The BCS-BEC crossover in an electron-hole bilayer system with unequal 
electron and hole densities was recently studied in Ref.~\onlinecite{pieri07} 
within a BCS mean field approach. 
Sarma and FFLO phases were found to be stable in some range of densities; electron-hole
bilayers appear thus promising candidates for the detection of such elusive phases.

In this paper we extend the work of Ref.~\onlinecite{pieri07} by including 
the in-plane Coulomb interactions, that were neglected there, as well as 
some screening effects.
We find that the effect of intra-layer 
Fock energy quantitatively changes the phase diagram moving the 
normal-condensed phase boundary to lower densities.
Comparing energy of the condensed phase 
with that of the normal phase, we map out the phase diagram in
the average density-population polarization plane. We check the
``local'' stability of the Sarma phase with respect to competing FFLO order 
by calculating the superfluid mass density and identify a negative superfluid 
mass density with an instability towards an FFLO phase. 
We calculate also the compressibility matrix in order to investigate 
possible instabilities towards phase separation.

Finally, we consider the effect of gate layer screening,
which proves especially important in the discussion of the local stability of the Sarma phase.
At zero temperature, the simultaneous presence of the singularity in the Coulomb potential and of a 
sharp Fermi surface produces in fact a logarithmic divergence in the 
momentum dependence of the BCS gap function which makes the Sarma phase always locally stable
against the FFLO phase. 
The inclusion of some form of screening removes this peculiar behavior, thus recovering the instability 
towards the FFLO phase in some region of the phase diagram.
 The intra-layer interactions and 
screening effects give therefore rise to a rich phase diagram in the 
crossover region between the BCS-like
high density state and the BEC of low density excitons, showing the
possibility to observe exotic superfluid phases as the population
polarization is changed.

The rest of this paper is organized as follows. In the next
section we outline the mean-field theory for electron-hole
bilayers and provide the set of self-consistent equations
for the quasiparticle energies and gap function. In Section
III, after a brief remark about our computational procedure, we
present our results for the quasiparticle properties and phase
diagram of the system. We conclude in Section IV with a summary
and outlook.

\section{Mean-Field Theory\label{mft}}
The Hamiltonian describing electrons and holes in a bilayer system
interacting with the Coulomb potential can be written as
\begin{widetext}
\begin{equation}
\begin{split}
\label{eq:ham}
\op{H}&=
\sum_\bk (\epsilon^a_\bk \ad_\bk a_\bk + \epsilon^b_\bk \bd_\bk b_\bk)
+\frac{1}{2A}\! \sum_{\bk_1\bk_2\bq }\!\! U^{aa}_\bq \ad_{\bk_1+\bq}
\ad_{\bk_2-\bq} a_{\bk_2} a_{\bk_1} \\
&\hfill+\frac{1}{2A} \sum_{\bk_1\bk_2\bq }\!\! U^{bb}_\bq \bd_{\bk_1+\bq}
\bd_{\bk_2-\bq} b_{\bk_2} b_{\bk_1}
+\frac{1}{A} \sum_{\bk_1\bk_2\bq }\!\! U^{ab}_\bq \ad_{\bk_1+\bq}
\bd_{\bk_2-\bq} b_{\bk_2} a_{\bk_1}
\end{split}
\end{equation}
\end{widetext}
The basis
states for electrons and holes are chosen to be plane wave states
labeled by two-dimensional wave vectors $\bk$ as is 
conventional for a uniform system.
The operators $a_\bk/\ad_\bk$ ($b_\bk/\bd_\bk$) are 
creation/annihilation operators for electrons (holes) respectively.
The single particle energies are denoted by
$\epsilon^a_\bk,\epsilon^b_\bk$ and the matrix element $U_\bq$ with 
respect to plane wave states becomes 
the Fourier transform of the corresponding two-body Coulomb interaction 
$U(\br)$
\begin{equation}
U^{aa}_\bq = U^{bb}_\bq = \frac{2\pi e^2}{\varepsilon q}\, ,    
\qquad 
U^{ab}_\bq = \frac{2\pi e^2}{\varepsilon q}    
\ee^{-qd}
\end{equation}  
where $U^{aa}, U^{bb}$, and $U^{ab}$ denote the electron-electron,
hole-hole and electron-hole Coulomb interactions, respectively, $A$
is the area of a layer and $d$ is the inter-layer separation. We disregard 
the spin degrees of freedom. 

The bilayer system is characterized by the electron and hole densities,
or equivalently by the average density parameter $r_s$ (average
distance between particles in the plane in units of Bohr radius 
$a_B$), the population polarization $\alpha$ (characterizing 
population imbalance in terms of the ratio of density difference and 
total density) defined by
\begin{equation}
n=\frac{1}{2}(n_a+n_b)=\frac{1}{\pi a_B^2 r_s^2} \quad \text{and} \quad
\alpha=\frac{n_a-n_b}{n_a+n_b} 
\end{equation}
and the inter-layer separation $d$.

The solution of the mean-field Hamiltonian at zero temperature ($T=0$) is given by 
the following coupled integral equations
\begin{eqnarray}
\Delta_\bk &=& - \frac{1}{A} \sum_{\bkp\neq \bk} U^{ab}_{\bk-\bkp}
\frac{\Delta_\bkp}{2E_\bkp} (1-f^+_\bkp-f^-_\bkp)\\
\xi_\bk &=& \epsilon_\bk-\mu-\frac{1}{2A}\sum_{\bkp\neq \bk} U^{aa}_{\bk-\bkp} 
[ (1-\xi_\bkp/E_\bkp)\nonumber\\
&&\phantom{aaaaaaa}\times(1-f^+_\bkp-f^-_\bkp)+f^+_\bkp+f^-_\bkp ]\label{csik} \\
E_\bk^2    &=& \xi_\bk^2 + \Delta_\bk^2 \\
f^{\pm}_\bk&=& \left\{ 
\begin{array}{lll}
1 & \textrm{if} & E^{\pm}_\bk < 0 \\
  &                            \\
0 & \textrm{if} & E^{\pm}_\bk > 0 
\end{array}
\right.
\end{eqnarray}
where $\epsilon_\bk=(\epsilon_\bk^a+\epsilon_\bk^b)/2$ (with $\epsilon_\bk^i=\hbar^2 k^2/2m_i$, $i=a,b$), 
the mean chemical potential $\mu=(\mu_a+\mu_b)/2$, while
\begin{eqnarray}
E^{\pm}_\bk&=&E_\bk \pm \Delta E_\bk\\
\Delta E_\bk&=&\Delta
\xi_\bk+\frac{1}{2A} \sum_{\bkp\neq \bk} U^{aa}_{\bk-\bkp} (f^-_\bkp-f^+_\bkp)\\
 \Delta \xi_\bk&=&\frac{1}{2}(\epsilon_\bk^a-\mu_a-\epsilon_\bk^b+\mu_b) .
\end{eqnarray}  
At finite temperature, the occupation functions
$f^\pm_\bk(E^\pm_\bk)$ go from the step function to the Fermi-Dirac 
distribution.

Given the electron and hole chemical potentials $\mu_a$ and $\mu_b$, 
these equations can be solved numerically to obtain the unknown functions
$\Delta_\bk, \xi_\bk$ and $\Delta E_\bk$. 
Physically, $\Delta_\bk$ is the BCS (s-wave) gap function while
$E^\pm_\bk$ are the quasi-particle excitation energies in the superfluid 
phase. In the absence of intra-layer interaction $\xi_\bk$ is 
just the average of the free electron and hole dispersions 
(with respect to the corresponding 
chemical potentials). Intra-layer interaction modifies the free 
dispersions by the inclusion of the exchange (Fock) interaction, 
as explicitly considered in Eq.~(\ref{csik}). 

For fixed number of particles the chemical potential values are 
adjusted to satisfy the number equations
\begin{equation}    
n_a =  \frac{1}{2A} \sum_{\bk}
       \left[ 
       \left(1+\frac{\xi_\bk}{E_\bk}\right) f^+_\bk +
       \left(1-\frac{\xi_\bk}{E_\bk}\right) (1-f^-_\bk) \right]
\end{equation}  
and
\begin{equation}    
n_b =  \frac{1}{2A} \sum_{\bk}
       \left[ 
       \left(1+\frac{\xi_\bk}{E_\bk}\right) f^-_\bk +
       \left(1-\frac{\xi_\bk}{E_\bk}\right) (1-f^+_\bk) \right] \, .
\end{equation}  
In the above mean-field description of the electron-hole
bilayer we have used the bare Coulomb interaction given in
Eq.\,(2). 
In realistic systems, the interactions 
entering the model  hamiltonian of Eq.\,(1) should be modified
to include
many-body effects such as exchange and correlation 
and
external potentials. These effects are described by a
screening function which usually decreases the strength of the
bare Coulomb interaction for electrons and holes in the normal
phase. However, the 2D screening due to intra- and inter-layer 
interactions is difficult to take into account properly for 
the condensed phase.\cite{gortel}
In order to see the qualitative effects of screening we
consider the mechanism of gate screening which can be taken into account 
in a simple way. In this mechanism the
Coulomb potential of a point charge is replaced by that of a dipole
consisting of the point charge and its image behind the
metallic gate. 
We have approximately modelled the screening by the gate potential 
by taking the 
intra- and inter-layer interactions to be
\begin{equation}
U^{aa}_\bq = U^{bb}_\bq = \frac{2\pi
e^2}{\varepsilon\sqrt{q^2+\kappa^2}}\, ,    
\qquad 
U^{ab}_\bq = \frac{2\pi e^2}{\varepsilon\sqrt{q^2+\kappa^2}}    
\ee^{-qd}\,
,
\end{equation}  
respectively, where the parameter $\kappa$ is a screening
wave number. In recent experiments with metallic gates to
control the charge densities, the separation between the gate
and 2D layer is about $\sim 250$\,nm \cite{seamons_apl,
screening}. Thus, the image charge is $\sim 500$\,nm away from 
the real charge and we may assume that for distances larger than
500\,nm, the Coulomb potential will be screened. In
our calculations we take the screening length associated with  
gate screening to be $\sim 20 a_B$, i.e., 
$\kappa = 1/20 a_B$. We have also checked other values of
$\kappa$ and found that the results are largely insensitive
in the range $40> 1/(\kappa a_B) >5$.

\section{Results and Discussion\label{results}}

\subsection{Numerical Procedure}
We solve the gap equations by representing the
unknown functions on a grid of $k$-points (after angular integration) 
and using a non-linear root finding scheme for the function values 
on grid points. For balanced populations an iterative scheme provides a
robust method of solution. 
For imbalanced populations we employ a root finding 
scheme for the function values on grid points and chemical potential 
values. We start with the equal density solution at the same 
average density and create imbalance
first at a small finite temperature and then decrease the temperature
(using the solution from the previous step as input) until results do
not change with temperature any more. 
The integrals are evaluated using
Gaussian quadrature. The finite temperature is necessary to obtain
smooth functions and gradients for the Newton-Raphson root finding
algorithm. We found it  necessary to introduce up to three different 
grids for integration to handle "discontinuities" at low temperatures, 
when one type of occupation number becomes equal to unity 
$f^\lambda_\bk=1$ ($\lambda=+$ or $-$) in a region of $k$-space 
(and causing the integrands to vanish there). 

The so-called Sarma states
obtained in this way are of the following BCS form
\begin{eqnarray}
\vert \Psi \rangle=\prod_{\bq \in \mathcal{R}} a^\dag_\bq
           \prod_{\bk \notin \mathcal{R}} (u_\bk+v_\bk a^\dag_\bk b^\dag_{-\bk})\vert 0 \rangle 
\end{eqnarray}
where the resulting wave function has a certain
range of $\bk$ states (the set denoted by $\mathcal{R}$) occupied with 
quasi-particles of the BCS theory giving rise to population imbalance.
The region $\mathcal{R}$ is where the quasi-particle energy 
$E^\pm_\bk$ becomes negative, i.e. less than that of the ground pair 
energy and the corresponding quasi-particle occupation becomes unity. 
Incidentally, the quasi-particles of BCS theory are just electron or 
hole states at that wave vector $\bk$.
Outside the set $\mathcal{R}$ we have pairs $\bk,-\bk$ of electrons and
holes.\cite{footnote}
Therefore, at $T=0$ there can be one or two Fermi surfaces depending 
where the set of $\bk \in \mathcal{R}$ 
vectors are. These topologically different phases 
will be called Sarma-1 (S1) and Sarma-2 (S2). These states have also been
called breached pair states~\cite{for05}
displaying a Fermi surface together with a condensate.
The gap function is non-zero but there are gapless excitations. 

\subsection{Quasi-Particle Properties}
In the following we present physical quantities in Rydberg 
units, i.e. length is measured in effective (excitonic) Bohr radius 
$a_B=\frac{\hbar^2\varepsilon}{me^2}$, momentum in $1/a_B$
and energy in effective Rydberg 
(Ryd= $\frac{\hbar^2}{2m a_B^2}=\frac{e^2}{2\varepsilon a_B}$). 
The reduced mass $m$ is defined by $1/m=1/m_a+1/m_b$
where $m_a=m_e$ and $m_b=m_h$ are the band mass of the electron and
hole, respectively. In the numerical calculations we specialize
to GaAs system parameters with mass ratio $m_a/m_b=0.07/0.30$ and
background dielectric constant $\varepsilon=12.9$.

\begin{figure*}[t]
\includegraphics[scale=1.0]{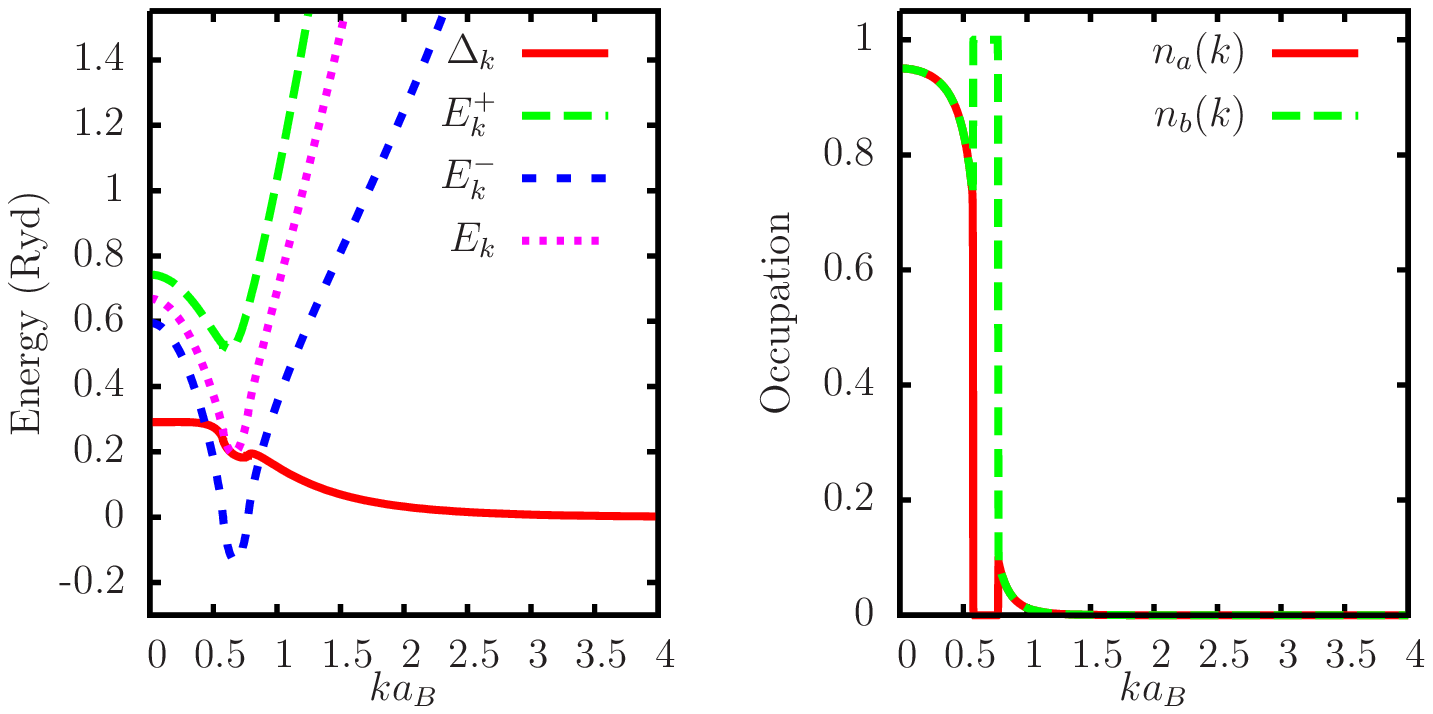}

\vspace{0.3cm}

\includegraphics[scale=1.0]{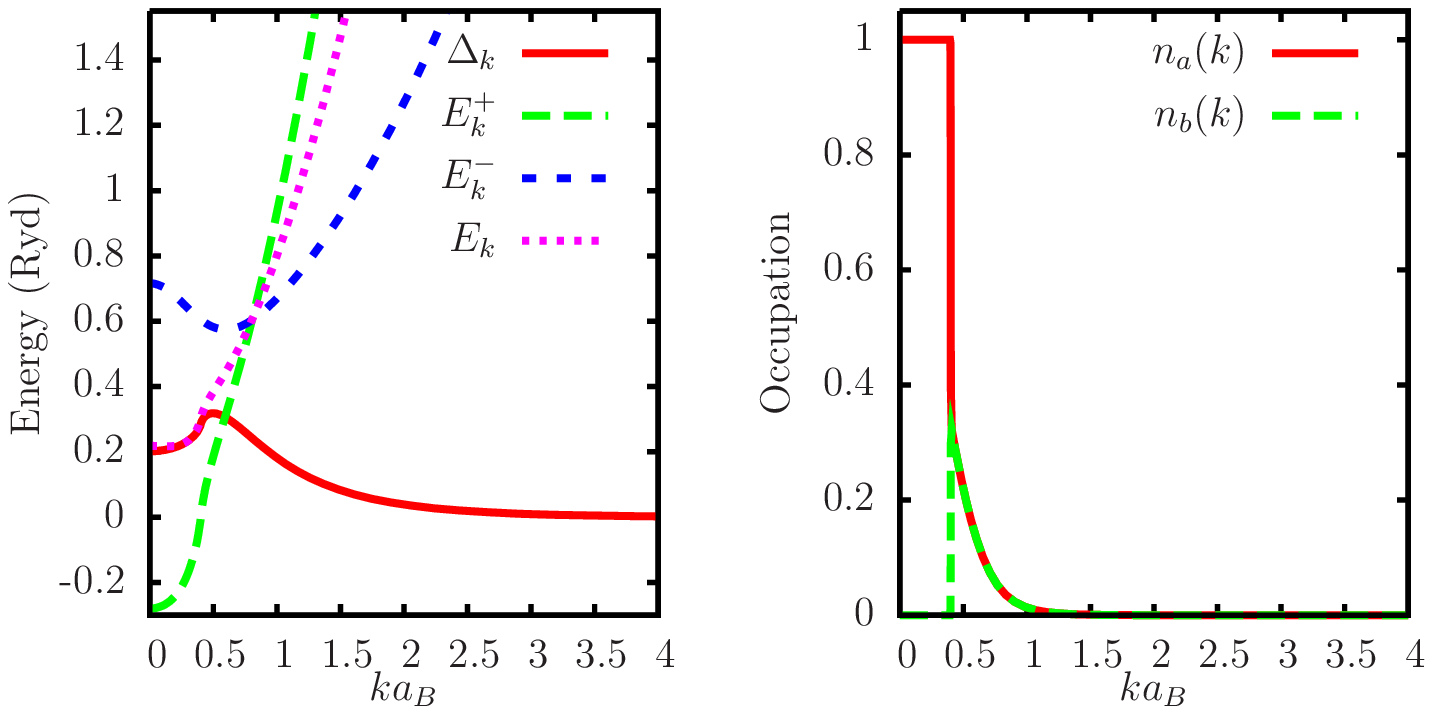}
\caption{(color online) Gap function and quasi-particle energies 
with bare Coulomb interactions for $m_e/m_h=0.07/0.30$ and $d=a_B$.
The upper panel shows a Sarma-2 phase at $r_s=3$ and $\alpha=-0.3$ 
with excess holes (heavy majority species).
The lower panel shows a Sarma-1 phase at $r_s=5$ and $\alpha=0.5$ 
with excess electrons (light majority species).
Occupation numbers are shown on the right.
\label{fig:gap}}
\end{figure*}

\begin{figure*}[t]
\includegraphics[scale=1.0]{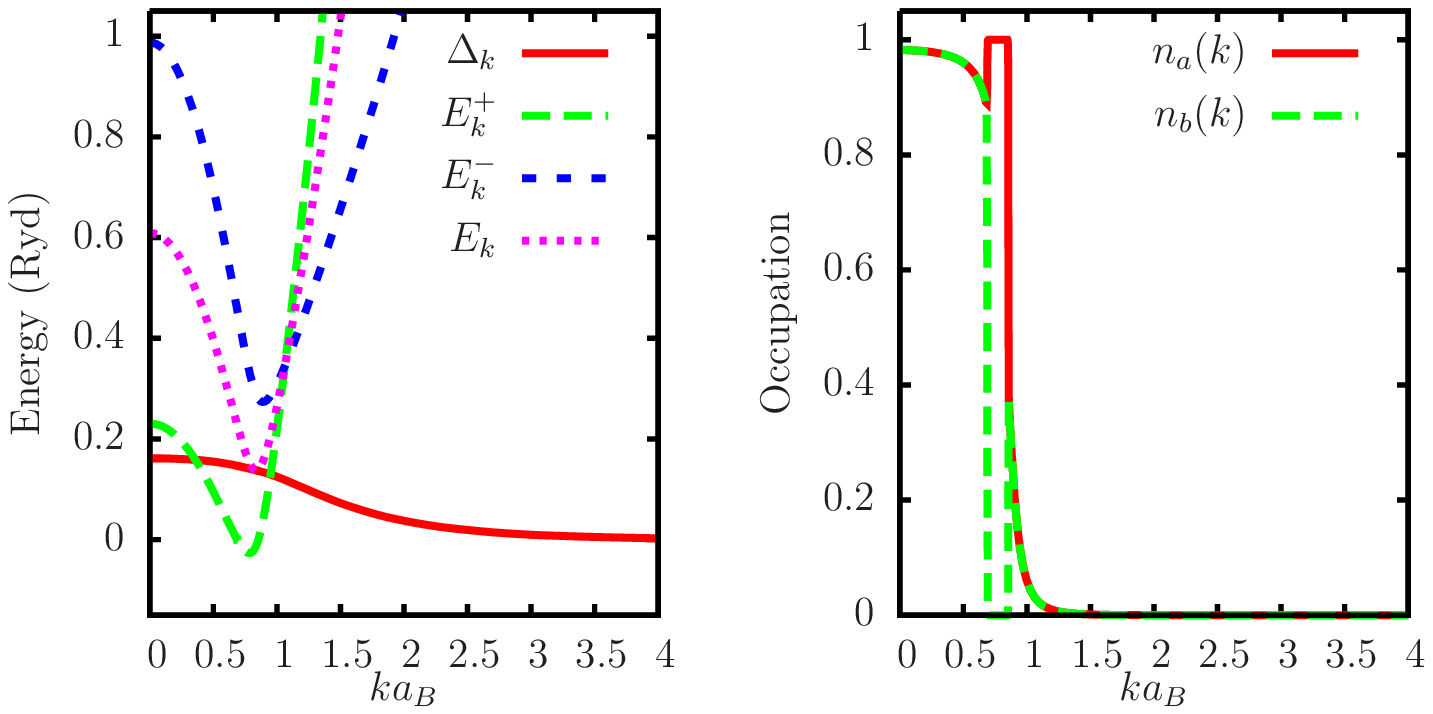}

\vspace{0.3cm}

\includegraphics[scale=1.0]{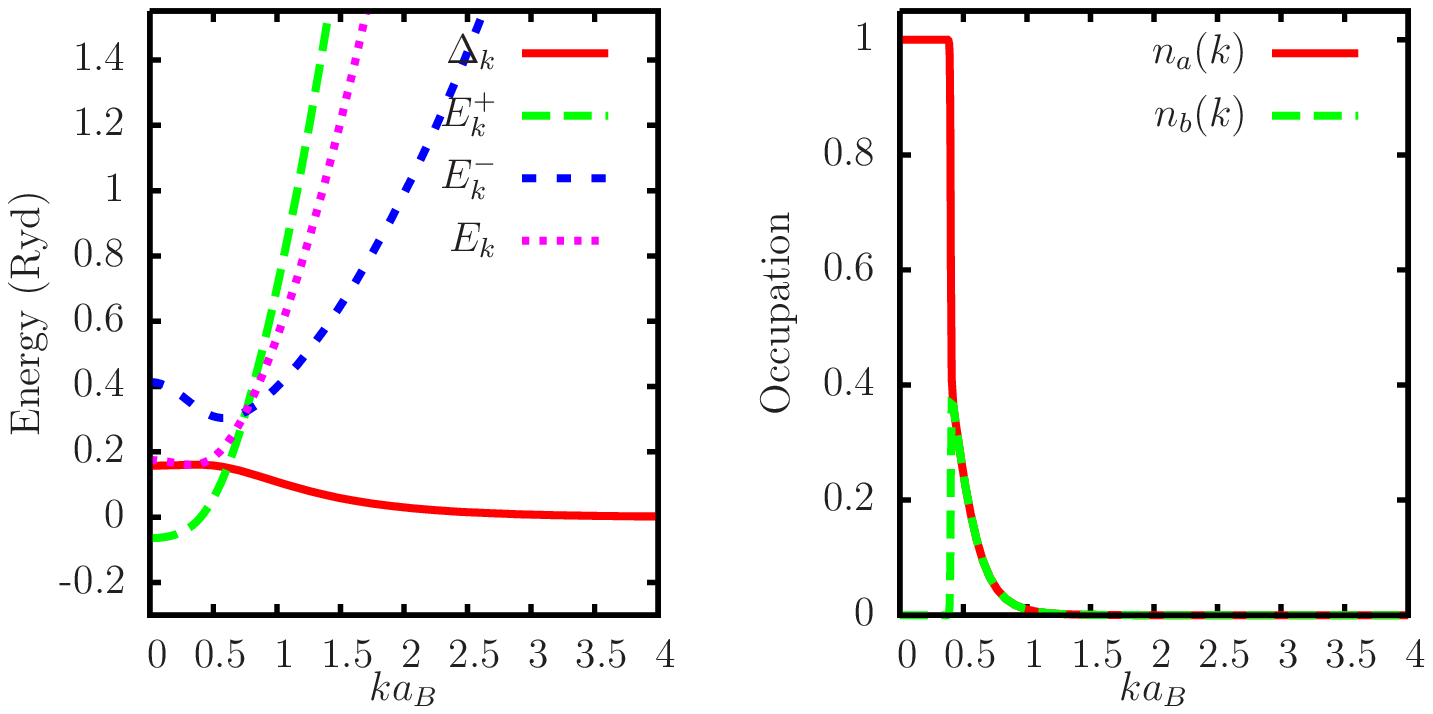}
\caption{(color online) Gap function and quasi-particle 
energies with screened Coulomb interactions for 
$m_e/m_h=0.07/0.30$ and $d=a_B$.
The upper panel shows a Sarma-2 phase at $r_s=2.5$ and $\alpha=0.2$ 
with excess electrons.
The lower panel show a Sarma-1 phase at $r_s=5$ and $\alpha=0.5$ 
with excess electrons.
Occupation numbers are shown on the right. The gap function has less
variation and the divergence in the derivative at the zero crossings
disappears.
\label{fig:gapscr}}
\end{figure*}

Representative solutions with one and two Fermi surfaces at $T=0$ are
illustrated in Fig.~\ref{fig:gap} (bare Coulomb interaction) and
Fig.~\ref{fig:gapscr} (screened interactions) for $d=a_B$. The figures show the gap
function $\Delta_\bk$, the quasi-particle energies $E^\pm_\bk$ 
and their average $E_\bk$ on the left panels, and the electron and 
hole occupation numbers 
$n_a(k), n_b(k)$ on the right panels. At $T=0$ in the ground state, the quasi-particle levels 
with negative energy are occupied, positive energy levels are empty.
The two different type of excitation branches 
are split both due to different electron-hole mass and chemical potential 
values. When one of the spectra crosses the zero energy axis, population 
imbalance is created.  
If the negative energy region includes the origin at $k=0$, the ground 
state has one Fermi surface, otherwise it has two. The two cases are 
denoted by S1 and S2 respectively.
The top panel in each figure shows an S2 phase and the bottom
panel shows an S1 phase.
Since the quasi-particle energy branch is continuous the system still 
has gapless excitations. A close investigation of the gap function 
$\Delta_\bk$  in the absence of screening (Fig.~\ref{fig:gap})
shows that it has a cusp at the zero crossings of the quasi-particle energy,
corresponding to a divergence in the derivative of  $\Delta_\bk$. This 
divergence has important consequences on the stability of the Sarma 
phase at $T=0$, as discussed  below.

\subsection{Superfluid mass density, compressibility matrix 
and the stability of the Sarma phase \label{stability}}

The ``local'' stability of the Sarma phase with respect to phases 
of the FFLO type is usually assessed by 
calculating the superfluid mass density 
(phase stiffness).\cite{pieri07,wu03,isk06} This quantity 
should be positive in a stable state and a negative value is 
identified with an instability towards an FFLO 
phase.\cite{ff,lo,pieri07} Clearly, the positivity of the superfluid mass
density guarantees only that the Sarma phase is a local minumum of the
energy with respect to fluctuations of the gap parameter associated
with pairing of the FFLO type, and does not exclude the possibility that an 
FFLO phase with finite pair momentum can be a global mimimum of the energy. 
When this happens, the local stability of the Sarma phase actually 
corresponds to metastability.   

\begin{figure*}[t]
\begin{center}
\includegraphics[scale=1.00]{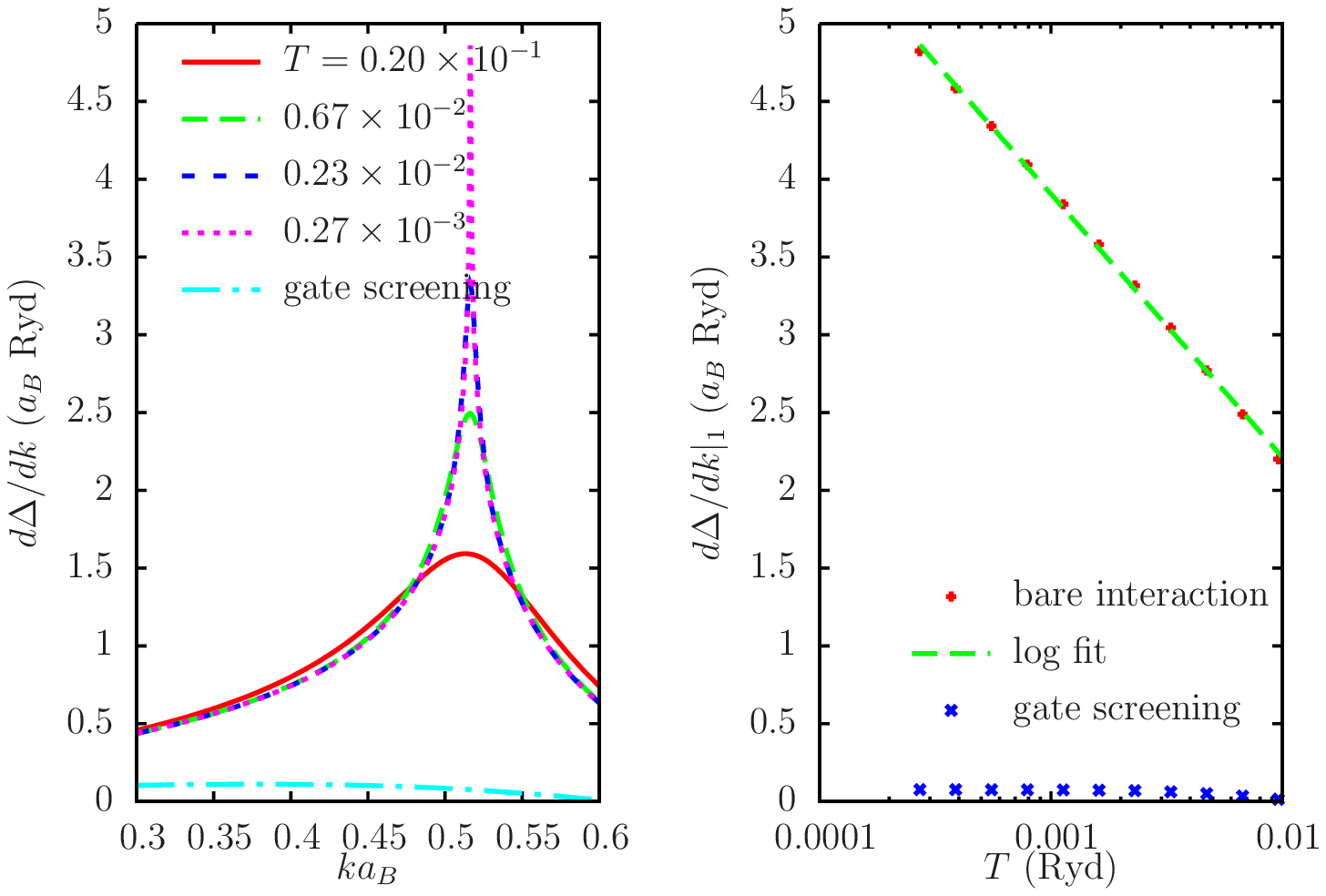}
\caption{The derivative of the gap function $d\Delta_\bk/dk$ as a 
function of $k$ at various temperatures with bare interlayer 
interactions, or at $T=0$ with gate screening (left). The value 
$d\Delta_\bk/dk|_1$ at the zero crossing of $E^+_\bk$ as a function 
of temperature $T$ with bare interlayer interactions or with gate 
screening (right). In both 
panels: $d=a_B,\,r_s=3,\,\alpha=0.3$ and 
$m_e/m_h=0.07/0.30$. 
\label{fig:deldiv}}
\end{center}
\end{figure*}

The superfluid mass density is given by~\cite{isk06}
\begin{eqnarray}
&&\rho_s = m_\mathrm{e} n_\mathrm{e} + m_\mathrm{h} n_\mathrm{h}
\nonumber\\
&&-
\frac{\hbar^2 \beta}{8 \pi} \int \dd k k^3
\frac{1}{2} \left[ \frac{1}{\cosh^2(\beta E^+_\bk/2)}
+
\frac{1}{\cosh^2(\beta E^-_\bk/2)}
\right]\phantom{aa}   
\end{eqnarray}  
where $\beta$ is the inverse temperature. At $T=0$ this expression 
can be written as~\cite{pieri07}
\begin{equation}
\rho_{s} \, = \, m_{\mathrm{e}} n_{\mathrm{e}} \, + \,
m_{\mathrm{h}} n_{\mathrm{h}} \, - \, \frac{\hbar^2}{4 \pi} \,
\sum_{j,\lambda} \, \frac{(k_{j}^{\lambda})^{3}}{\left|
\frac{dE^{\lambda}_k}{dk} \right|_{k=k_{j}^{\lambda}}} \, .
\label{superfluid-density}
\end{equation}
where $k_{j}^{\lambda}$ are the roots of $E^{\lambda}_k$ with
$\lambda=\pm$.
At zero temperature the last expression involves the derivative 
of $\Delta_k$ (through the derivative of $E_k^\pm$) at the zero 
crossings of $E_k^\pm$.
As mentioned above, our calculations are carried out at nonzero but 
small temperature. 
We have found that this derivative diverges logarithmically as $T\to 0$. 
In particular, we have demonstrateed analytically that for the bare 
Coulomb interaction one has
\begin{equation}    
\left| \frac{ \dd \Delta_\bk}{\dd k} \right|_{k=k^*} \approx
\frac{e^2}{\pi \varepsilon} 
\frac{\Delta_{\bk^*}}{2 E_{\bk^*}} 
\left|\ln T \right| \quad \textnormal{as} \quad T \to 0
\end{equation}
where $k^*$ is the zero crossing point  at $T=0$ as $k \to k^*$
\begin{equation}    
\left|\frac{ \dd \Delta_\bk}{\dd k}\right|_{T=0} \approx
\frac{e^2}{\pi \varepsilon} 
\frac{\Delta_{\bk^*}}{2 E_{\bk^*}} 
\ln \left|k-k^*\right| \quad \textnormal{as} \quad k \to k^*,
\end{equation}
which we have also checked numerically (Fig.~\ref{fig:deldiv}).
This divergence is due to the simultaneous presence of the long range 
Coulomb interaction, which is singular at $q=0$, and the 
discontinuity of the Fermi function at $T=0$. 
Finite temperature and/or screening effects, smear out these 
singularities thus removing the divergence.
The presence of this divergence at $T=0$ was overlooked in the 
previous mean field study of the 
imbalanced electron-hole bilayer system~\cite{pieri07}. 
As a matter of fact, making this divergence emerge from the numerical 
calculation requires the achievement of very low temperatures in 
the calculations and an extreme precision 
in the numerical integration.
  
The presence of this divergence is particularly meaningful 
for the analysis of the local stability of the Sarma phase at 
strictly $T=0$. The diverging derivative makes in fact the 
negative contribution to $\rho_s$ vanish, thus implying 
that for the unscreened Coulomb interaction the Sarma phase is 
always locally stable at $T=0$.
This result is interesting as a matter of principle, as it 
offers an ``extreme'' example, where the argument by 
Forbes {\it et al}. \cite{for05} 
that mass ratio and momentum structure of the interactions should 
favor the stability of Sarma phase is completely effective.
On the other hand we expect it to have few practical consequences,
since the stability of the Sarma phase 
induced by this divergence is very fragile with 
respect to finite temperature and/or screening effects.
As Fig.~\ref{fig:deldiv} clearly shows, quite small temperatures  
suffice to smear out the divergence in the derivative.
Alternatively, the simple screened interaction makes the divergence
to disapper even at $T=0$, as also shown
in Fig.~\ref{fig:deldiv}.
As a result, in the presence of screening, the Sarma phase
will indeed be locally unstable in certain regions of the 
$r_s-\alpha$ plane, as discussed in the next subsection. 

The mechanical stability of the system with respect to phase 
separation requires the compressibility matrix 
$\partial \mu_i/\partial N_j$ to be positive definite.
We have therefore calculated the compressibility matrix 
across our phase diagram to check also this
stability. When the intra-layer Coulomb interaction is neglected, 
the compressibility matrix develops
negative eigenvalues across most of our phase diagram (restricting 
the stable region only to small values of $r_s$) in agreement 
with the findings of the recent work by 
Yamashita {\it et al.}\cite{ohashi09}.
However, as it was already argued in Ref.~\onlinecite{pieri07}, 
this apparent dominant instability towards phase separation is 
an artifact occurring when the intra-layer Coulomb repulsion 
is artificially excluded from the calculation. 
It should therefore not be taken seriously. 
In particular, we have verified explicitly that in our calculations
{\em with\/} Coulomb intra-layer repulsion, 
the Hartree term, which increases linearly with the distance 
$d_G$ between the metallic gates and the electron/hole layers, washes out 
completely phase separation from our phase diagram of 
Fig.~\ref{fig:phased1} already for distances $d_G$ of the order 
of 5-10 $a_B$, well below the typical 
gate-to-layer distances in current devices. 
We thus conclude that, contrary to what happens in cold atom 
systems, phase separation is not an issue in
electron-hole bilayer systems.

\subsection{Phase Diagram at $d=a_B$ \label{phasediagram}}

\begin{figure*}[t]
\begin{center}
\includegraphics[scale=1.0]{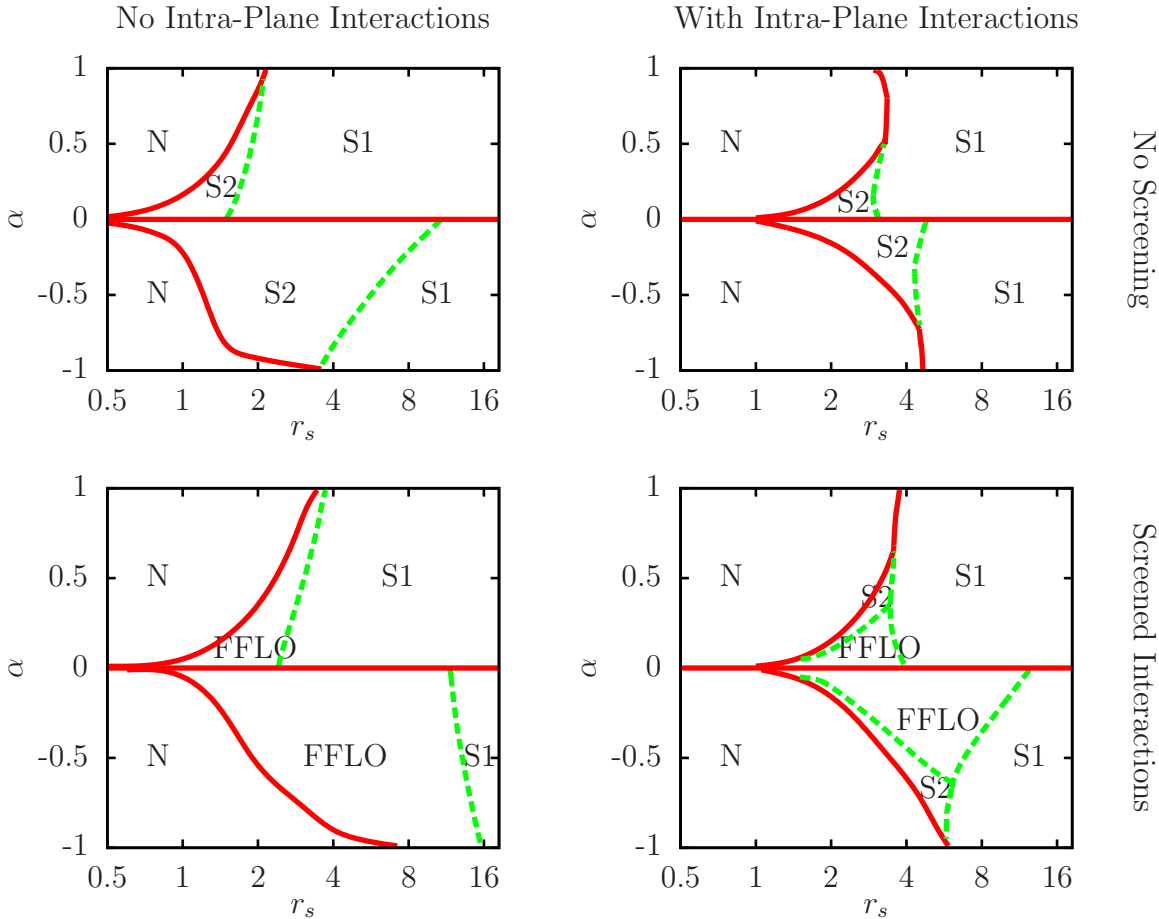}
\caption{(color online) Phase diagram for inter-layer separation 
$d=a_B$ with progressive refinement of the approximations. Superfluid (S1/S2)-normal (N) 
phase boundaries are shown 
with red solid lines. A negative
superfluid mass density showing a local instability is assumed to be
towards an FFLO phase. S1, S2 and FFLO boundaries are shown with green
dashed lines. The four panels corresponds to calculations including: bare inter-layer interactions
only (upper left), bare intra- and inter-layer interaction (upper
right), gate screened inter-layer interactions only (lower left) and
gate screened intra- and inter-layer interaction (lower right).
The $\alpha=0$ line is special in the phase diagram and corresponds to the BCS state with 
equal populations, which is always stable. 
\label{fig:phased1}}
\end{center}
\end{figure*}

In this section we present the phase diagram resulting from the comparison of 
the energies of the Sarma and normal phases and from the stability analysis 
discussed in the previous section.
We set the inter-layer separation equal to one effective
(excitonic) Bohr radius $d=a_B$. 
To make contact with previous literature, we present in Fig.~\ref{fig:phased1}
the phase diagram for 
progressively refined approximations corresponding to the inclusion in the
calculations of: (i) bare inter-layer interactions only, (ii) bare
inter- and intra-layer interactions, (iii) screened inter-layer
interactions only and (iv) screened inter- and intra-layer interactions.


For bare interactions, the superfluid density is always positive and the Sarma 
phase is ``locally'' stable, due to the mechanism explained in 
Sec. \ref{stability}. Therefore, in the top
panel of Fig.~\ref{fig:phased1} we do not show any FFLO phase, but our
calculations do not rule out the possibility of a first order transition to an
FFLO phase with a finite FFLO modulation momentum ${\bf q}$ as found in 
Ref.~\onlinecite{ohashi09}.
Two topologically distinct Sarma phases, Sarma-1 with
one Fermi surface and Sarma-2 with two Fermi surfaces, are present in
the phase diagrams. The effect of the intra-layer repulsive interactions is to 
favor the normal phase with respect to the condensed phases, thus shifting 
to higher vaues of $r_s$ the boundary between normal and condensed phases 
(right panels). 
The two bottom panels of Fig.~\ref{fig:phased1} presents the phase diagram when
the gate screening is taken into account. With inter-layer interactions only, 
the Sarma phase becomes unstable for a large portion of the phase diagram,
especially with excess holes, i.e. $\alpha<0$. (bottom left panel) There
is no stable S2 phase in this case. Switching on the intra-layer interactions
reduces the space occupied by the FFLO phase in our phase diagram and 
restores the S2 phase in some region of the phase diagram.
This result is physically quite sensible, as the FFLO modulations 
of the gap parameter should be unavoidably accompanied by some 
modulations of the density in real space.
In the presence of the Coulomb intra-layer repulsion such 
density modulations are energetically expensive, thus hindering 
the FFLO phase with respect to the Sarma phase.\cite{footnote2}

A quite rich phase diagram is therefore obtained when
both intra-layer and screening effects are present. The presence of 
locally stable Sarma phases confirms the expectation that isotropic 
translationally invariant gapless superfluid states can be stable with 
momentum dependent interaction.\cite{for05}
We note in this context that in recent work, Sarma phases were found to be 
stable  also in two-dimensional, two-band neutral Fermi systems.\cite{he09,des09}

\section{Summary\label{summary}}

We have studied a bilayer system of electron and hole
layers spatially separated by an insulating barrier, where
the electron and hole densities can be controlled independently and
have analyzed $s$-wave pairing between electrons and holes as a function of
average density and population difference using mean-field theory. 
By solving the
relevant energy gap equations we have compared the energy of the 
condensed phase called the Sarma phase with that of the normal state 
which is the sum of the electron and hole Fermi liquid energy 
described by the Hartree-Fock solution. 
We have included both inter- and intra-layer interactions
generalizing earlier work which did not include in-plane
interactions.~\cite{pieri07} 
In this way the phase boundary for the ground state is
established in the density-population polarization ($r_s-\alpha$) 
plane. The ``local'' stability of the Sarma phase was checked by 
calculating the superfluid mass density, whereas the
stability with respect to phase separation was assessed by calculating the
compressibility matrix.
We have found that with bare Coulomb interactions the Sarma
phase is always locally stable due to a peculiar momentum structure of the 
gap function originating from the singular infrared behavior of the Coulomb potential, and the simultaneous presence of a sharp Fermi surface at zero temperature.

Employing a simple model of screening which introduces an 
infrared cut-off in the Coulomb interaction, we have found 
that some regions in the phase space become unstable. 
We interpret this as an instability
towards an FFLO phase. Together with intra-layer interactions, the phase
diagram in the crossover regime from the weakly interacting high density
BCS limit to the strongly interacting BEC of dilute excitons has room
for various phases. The topologically different S1 and S2 Sarma phases
and FFLO are present with the inclusion of screening and intra-layer
interactions. On the other hand, without any screening there is no
instability towards FFLO and turning off intra-layer interactions 
the phase diagram does not show an S2 state. Currently, the 
experimental situation allows these systems to be 
realized.~\cite{seamons_prl,croxall_prl,cro09} 
Quantitative comparison would require a more realistic model of
screening, accounting for the condensed phase and finite width 
of the quantum wells, incorporating the disorder effects, 
and inclusion of spin degrees of freedom which may enter non-trivially 
when there are spin dependent interactions such as spin-orbit coupling.
With the renewed mean-field phase diagram at hand,
it would also be interesting to perform QMC simulations to
probe the predicted phases. 

\acknowledgments{This work is supported by TUBITAK (No. 108T743), TUBA,
European Union 7th Framework project UNAM-REGPOT
(No. 203953), and the Italian MUR under Contract No.
PRIN-2007 ``Ultracold Atoms and Novel Quantum Phases''.
A.\,L.\,S. thanks the hospitality of the Theoretical Physics
Department of the University of Trieste during the time he spent
there at an earlier stage of this work through a 
TUBITAK-BAYG scholarship.}


\begin{thebibliography}{99}
\bibitem{footnote}
For QMC applications this state can be written 
as a Slater determinant in real space.
\bibitem{footnote2} 
For the same reason we expect that inclusion of intra-layer 
interaction could  strongly affect the mean-field phase diagram of 
Ref.~\onlinecite{ohashi09}.
\end{thebibliography}
\end{document}